\documentstyle[prd,aps]{revtex}
\begin{document}
\draft
%%%%%%%%%%%%%%%%%%%%%%%%%%%%%%%%%%%%%%%%%%%%%%%%%%%%%%%%%%%%%%%%%%
%
%  Uncomment following two lines and one below for 2 column format.
%
\twocolumn[\hsize\textwidth\columnwidth\hsize\csname
@twocolumnfalse\endcsname
%%%%%%%%%%%%%%%%%%%%%%%%%%%%%%%%%%%%%%%%%%%%%%%%%%%%%%%%%%%%%%%%%%
\title{A Comment on the Estimation of Angular Power Spectra in the
Presence of Foregrounds}
\author{Martin White}
\address{Department of Astronomy and Department of Physics\\
University of Illinois at Urbana-Champaign\\
Urbana, IL 61801-3080}
\date{\today}
\maketitle
\begin{abstract}
It is common practice to estimate the errors on the angular power spectrum
which could be obtained by an experiment with a given angular resolution
and noise level.  Several authors have also addressed the question of
foreground subtraction using multi-frequency observations.
In such observations the angular resolution of the different frequency
channels is rarely the same. In this report we point out how the ``effective''
beam size and noise level change with $\ell$ in this case, and give an
expression for the error on the angular power spectrum as a function of $\ell$.
\end{abstract}

%  This is the other line to be uncommented for 2 column format
\vskip2pc]

\section{Introduction}

There are many experiments planned which hope to reap the rich harvest of
information available in the anisotropies of the Cosmic Microwave Background
(CMB) \cite{BenTurWhi}.
For the purposes of estimating cosmological parameters or constraining models
it is important to know how well an experiment can constrain the angular
power spectrum of CMB anisotropies.

It is common practice to estimate the errors on the angular power spectrum
which could be obtained by an experiment with a given angular resolution
and noise level.  Such an estimate can give us insight into what regions of
the angular power spectrum one could constrain, and what are the limiting
factors in an experiment designed e.g.~to constrain cosmological parameters.
In this report we point out some simple limiting cases which allow one to
gain intuition about the effect of multiple observing frequencies with
differing beam sizes in the presence of (somewhat idealized) foregrounds.

If we write the angular power spectrum of the anisotropy
as $C_\ell$ and the noise power spectrum as $N_\ell$
(typically $N_\ell=4\pi f_{\rm sky}\sigma^2/N_{\rm pix}$, where $f_{\rm sky}$
is the fraction of sky covered, $\sigma$ is the rms pixel noise and
$N_{\rm pix}$ is the number of pixels) then
\begin{equation}
\delta C_\ell \simeq \sqrt{ {2\over (2\ell+1) f_{\rm sky}} }
  \left( C_\ell + {N_\ell\over W_\ell} \right)
\label{eqn:dcl}
\end{equation}
where $W_\ell$ is the window function of the experiment and we have assumed
that the noise is gaussian.
For a gaussian beam of width $\theta_b$, $W_\ell=\exp(-\ell^2\theta_b^2)$.
Eq.~(\ref{eqn:dcl}) assumes that the only sources of variance in an experiment
are the anisotropy signal and the receiver noise, i.e.~it neglects foregrounds.
We have also implicitly assumed that $\delta C_\ell$ is a gaussian error,
thus we imagine we are working at reasonable high-$\ell$ and will bin our
power spectrum estimates into finite width bins in $\ell$.
Eq.~(\ref{eqn:dcl}) has been widely used to estimate how well upcoming
satellite experiments could constrain cosmological parameters \cite{Param}.

Several authors have also addressed the question of foreground subtraction
using multi-frequency observations \cite{Brandt,Dod,TegEfs,Teg}.  With
multi-frequency observations it appears possible to separate the desired
anisotropy signal from the foregrounds with encouraging precision.
The authors of \cite{Brandt,Dod} have shown that one can regard foreground
subtraction as an enhancement of the noise in a foreground free experiment.
The noise enhancement factor has been called the {\it foreground degradation
factor\/} (FDF) by Dodelson \cite{Dod} who gave a simple expression for it.
A method of foreground subtraction using both frequency and spatial information
was proposed in Ref.~\cite{TegEfs}.
Various methods of foreground subtraction have been compared recently by
Tegmark \cite{Teg}.

Note that the ``noise term'' in Eq.~(\ref{eqn:dcl}) depends exponentially
on $\ell$ once $\ell > \theta_b^{-1}$.
In multi-frequency observations the angular resolution of the different
frequency channels is rarely the same, leading to the question of which beam
size to use.
In this report we discuss a simple heuristic ``effective'' beam size and
noise level.
Since how well we can take out foregrounds depends on the angular structure
of the foregrounds, the ``noise'' will be a function of the foreground angular
power spectrum.

The formalism is directly comparable to that in \cite{TegEfs,Teg} in that
we use a minimum variance estimator of the CMB anisotropy power spectrum.
An extension to ``real world foregrounds'' \cite{Teg} is straightforward,
but the purpose of this report is to gain intuition through simple examples
so we do not pursue this line of development.

\section{Formalism}

As noted by Tegmark \cite{Teg}, one can consider foregrounds as an additional
noise component which is highly correlated among different frequency channels.
If we label the frequency channels by a greek subscript we can define a noise
correlation matrix
\begin{equation}
N^\ell_{\alpha\beta} = {4\pi\over N_{\rm pix}} \sum_i
  W_{\ell\alpha}^{1/2}
  \left\langle f^i_{\ell\alpha} f^i_{\ell\beta}\right\rangle
  W_{\ell\beta}^{1/2}
  + {4\pi\over N_{\rm pix}} \sigma^2_\alpha \delta_{\alpha\beta}
\end{equation}
which has contributions from the pixel noise (assumed uncorrelated between
channels here for simplicity) and foregrounds labelled by superscript $i$.
Here $\langle f^i_{\ell\alpha} f^i_{\ell\beta}\rangle$ is the
correlation matrix of foreground $i$ in channels $\alpha$ and $\beta$, with
the $4\pi/N_{\rm pix}$ inserted for later convenience.
If the foregrounds are 100\% correlated between the channels then
$\langle f^i_{\ell\alpha}f^i_{\ell\beta}\rangle=f^i_{\ell\alpha}f^i_{\ell\beta}$
where $f^i_{\ell\alpha}$ is the rms intensity as a function of frequency.
We shall assume this case from now on, but see Ref.~\cite{Teg}.

It is straightforward to derive the minimum variance estimator of $C_\ell$,
as a linear combination of measurements at different frequencies.  If we have
measured multipole moments $a^{\alpha}_{\ell m}$ at frequency $\alpha$ then
we write the estimate of the CMB component as
$\theta_{\ell m}=\sum_\alpha F_{\alpha} a^{\alpha}_{\ell m}$.
Imagine that we can write the observed signal
$a^{\alpha}_{\ell m}=t_{\ell m} W_{\ell\alpha}^{1/2}+n_{\ell m\alpha}$,
where $t_{\ell m}$ is the cosmological signal of interest (the same in each
frequency channel) and $n_{\ell m\alpha}$ is the sum of the noise and
foregrounds (i.e.~the non-CMB components).
We now minimize the variance of our estimator $\theta_{\ell m}$ minus
the ``real'' underlying sky ($t_{\ell m}$) with respect to the weighting
matrix $F_{\alpha}$.  We find that our minimum variance $C_\ell$ estimator
is (averaging over $m$)
\begin{equation}
\widehat{C}_\ell = \sum_{\alpha\beta}
  {\theta_{\ell}^2 - F_{\alpha} N^\ell_{\alpha\beta} F_{\beta} \over
   W^{1/2}_{\ell\alpha} F_{\alpha} \ W^{1/2}_{\ell\beta} F_{\beta} }
\label{eqn:clest}
\end{equation}
where $\theta_{\ell}^2$ is the average of $\theta_{\ell m}^2$ over $m$,
\begin{equation}
F_{\alpha} = \sum_\beta C_\ell W_{\ell\beta}^{1/2} \ \left(
  W_{\ell\beta}^{1/2} C_\ell W_{\ell\alpha}^{1/2} +
  N^{\ell}_{\alpha\beta} \right)^{-1}
\end{equation}
and we have left the $\ell$-dependence of $F_\alpha$ implicit for
notational convenience.  The vector $F_\alpha$ projects out the $\ell$th
CMB multipole moment from the signal in each channel in a minimum
variance sense.

If we assume that this estimator is Gaussian then we can replace
Eq.~(\ref{eqn:dcl}) by
\begin{equation}
\delta C_\ell = \sqrt{ {2\over (2\ell+1)} }
  \left( C_\ell + \sum_{\alpha\beta}
  {(F_{\alpha} N_{\alpha\beta} F_{\beta})_\ell \over
  (W_{\ell\alpha}^{1/2} F_{\alpha})^2 } \right)
\label{eqn:dcl2}
\end{equation}
where we have set $f_{\rm sky}=1$ for simplicity (the scaling with
$f_{\rm sky}$ is given in Eq.~(\ref{eqn:dcl})).
Note that in the limit of one frequency channel and no foregrounds, the
sums over $\alpha$ and $\beta$ are trivial, the noise term is independent
of $F$ and we recover Eq.~(\ref{eqn:dcl}).

Eq.~(\ref{eqn:dcl2}) is the general result, and we consider several examples
to gain intuition in the next section.

\section{Examples}

Let us consider various limits of Eq.~(\ref{eqn:dcl2}).  We will focus on
the noise term, since the part of the error proportional to $C_\ell$ simply
reflects cosmic plus sample variance \cite{ScoSreWhi}.
In the signal dominated limit ($C_\ell\gg N_\ell$) $F_\alpha$ is just the
inverse square root of $W_{\ell\alpha}$ and the noise term in
Eq.~(\ref{eqn:dcl2}) reduces to
\begin{equation}
\left( {N_\ell\over W_\ell} \right)_{\rm eff} =
  {4\pi\over N_{\rm pix}W_\ell^2} \left[
  \sum_\alpha \sigma_\alpha^2 W_{\ell\alpha} +
  \left(\sum_\alpha f_{\ell\alpha}W_{\ell\alpha}^{1/2}\right)^2 \right]
\label{eqn:signaldominated}
\end{equation}
where $W_\ell=\sum_\alpha W_{\ell\alpha}$.  Thus, in the absence of foregrounds
the noise is the sum of the noises in those channels able to resolve features
with multipole number $\ell$, divided by the number of such channels squared.
The addition of foregrounds increases the variance, but by assumption both
contributions to $\delta C_\ell$ are sub-dominant.
We expect this regime to occur at low-$\ell$, where we are cosmic and
sample variance limited.

Now let us consider the opposite limit.  First imagine there are no
foregrounds.
In this limit ($N_\ell\gg C_\ell$) the noise term in Eq.~(\ref{eqn:dcl2})
reduces to
\begin{equation}
\left( {N_\ell\over W_\ell} \right)^{-1}_{\rm eff} = 
  {N_{\rm pix}\over 4\pi} \sum_\alpha {W_{\ell\alpha}\over \sigma_\alpha^2 }
\label{eqn:noisedominated}
\end{equation}
which would be the obvious way to combine the channels to obtain the
effective noise: recall the error on a weighted mean is
$\sigma^{-2}=\sum_i \sigma_i^{-2}$.

Now we can enhance this last example by the addition of foregrounds.
For simplicity imagine a two channel experiment with 1 foreground
$f_{\ell\alpha}$ in addition to uncorrelated noise $\sigma_\alpha$,
with $\alpha=1,2$.
For definiteness imagine that channel 1 is a low frequency channel with a
``large'' beam, while channel 2 is a high frequency channel with a ``small''
beam.
Some trivial matrix algebra allows us to write $(N_\ell/W_\ell)_{\rm eff}$ as
\begin{equation}
{4\pi\over N_{\rm pix}}
\ {\sigma_1^2\sigma_2^2 + \sigma_1^2 f_{\ell 2}^2 W_{\ell 2} +
   \sigma_2^2 f_{\ell 1}^2 W_{\ell 1} \over
   \sigma_1^2 W_{\ell 2} + \sigma_2^2 W_{\ell1} +
   [f_{\ell 1}-f_{\ell 2}]^2 W_{\ell 1}W_{\ell 2} }
\end{equation}
which reduces to Eq.~(\ref{eqn:noisedominated}) as $f_\alpha\to0$.

To bring out the essential details let us take $\sigma_1=\sigma_2$ and
ignore the different resolutions by working at low enough $\ell$ that
$W_{\ell\alpha}\to1$,
\begin{equation}
\left( {N_\ell\over W_\ell} \right)_{\rm eff} 
\to {4\pi\sigma^2\over N_{\rm pix}}
\left( {\sigma^2 + f_{\ell 1}^2+f_{\ell 2}^2 \over
 2\sigma^2 + (f_{\ell 1}-f_{\ell 2})^2 } \right) \quad .
\label{eqn:fdfgeneral}
\end{equation}

The term in parentheses is the increase in the noise over the one channel
result.  As we decrease $\sigma^2$, this goes from ${1\over 2}$
(the case of no foregrounds: co-add the channels) to $(1+x^2)(1-x)^{-2}$ where
$x=f_{\ell 2}/f_{\ell 1}$.  For this case, the increase in the noise is just
the FDF defined by Dodelson \cite{Dod}, written in a slightly different
notation.
The point $x=1$ is when the frequency dependence of the foreground is the
same as the CMB.
Generically the minimum variance estimator Eq.~(\ref{eqn:clest}) does better
than the FDF would indicate, as has been pointed out in Ref.~\cite{Teg}.

Now let us reinstate the window functions, but imagine we are at intermediate
$\ell$, where $W_{\ell 1}\ll 1$.  In this regime
\begin{equation}
\left( {N_\ell\over W_\ell} \right)_{\rm eff} =
  {4\pi\sigma_2^2\over N_{\rm pix}W_{\ell 2}}
  \ \left( 1 + { f_{\ell 2}^2 W_{\ell 2}\over \sigma_2^2} \right) \quad .
\end{equation}
So the appropriate resolution is that of channel 2, and the noise is
the channel 2 noise {\it plus\/} a contribution from the foreground at the
higher frequency.  Note that the properties of channel 1 do not enter the
expression, as expected.
The foreground contribution declines as one approaches the resolution of
channel 2, so that asymptotically the error is just
$4\pi\sigma_2^2/N_{\rm pix}W_{\ell 2}$.

Finally note that if our foregrounds have a ``steep'' power spectrum,
i.e.~fall rapidly with $\ell$, then at high-$\ell$ we have the limit
$f_{\ell\alpha}\to0$ and reproduce Eq.~(\ref{eqn:noisedominated}).
At low-$\ell$ the foreground should dominate over the noise which is the
case discussed below Eq.~(\ref{eqn:fdfgeneral}).
This shows that for foregrounds without a lot of small-scale structure there
is little effect at high-$\ell$ from removing the foregrounds
(in this simplified example where the foreground properties are assumed known
exactly).

\section{Summary}

Let us summarize our results with $\ell$ increasing from the signal- to
the noise-dominated limits.
Under our assumptions, the error on the angular power spectrum starts
dominated by cosmic and sample variance at low-$\ell$.
Moving to higher $\ell$, if the noise or foregrounds start to dominate on
scales which all channels can resolve, the effect of a foreground looks like
an increase in the noise by a factor which is always less than Dodelson's
FDF \cite{Dod}.
If the noise or foregrounds dominate on scales smaller than the size of
the widest beam, but larger than the size of the smallest beam, the noise
is that of the channel with the smaller beam size, increased by the variance
of the foreground at that frequency.
At $\ell>\theta_b^{-1}$ for even the highest frequency channel the error
is simply that obtained from the noise in the highest frequency channel,
regardless of foregrounds.
If the foregrounds are always negligible compared to the noise, the
appropriate noise level is the weighted sum of the noises in each channel
which can resolve the given $\ell$, as shown in
Eq.~(\ref{eqn:noisedominated}).

\section*{Acknowledgments}

I thank Douglas Scott for useful comments on the manuscript.

\end{document}